\documentclass[11pt,a4paper,oneside]{article}

\usepackage{mymacro}
\usepackage{verbatim}
\usepackage{tikz}
\usetikzlibrary{arrows.meta}
 \usetikzlibrary{shapes.geometric}
\definecolor{myRed}{RGB}{150,22,22}
\protected\edef\mathcal{%
  \unexpanded\expandafter\expandafter\expandafter{%
    \csname mathcal \endcsname
  }%
}

\DeclareMathOperator{\dDisc}{dDisc}

\title{\textbf{Bubbles in AdS}}

\author{Agnese Bissi$^{a,b,c}$,}
\author{Giulia Fardelli$^d$}
\author{and Mohammad Reza Khansari$^{b,e}$}

\affiliation{$^{a}$ICTP, International Centre for Theoretical Physics, \\
				Strada Costiera 11, 34151, Trieste, Italy}
\affiliation{$^{b}$INFN, Sezione di Trieste, Via Valerio 2, 34127 Trieste, Italy,}
\affiliation{$^{c}$Department of Physics and Astronomy, Uppsala University,\\
Box 516, SE-751 20 Uppsala, Sweden}

\affiliation{$^{d}$Department of Physics, Boston University, Boston, MA 02215, USA}

\affiliation{$^{e}$SISSA, Via Bonomea 265, 34136 Trieste, Italy}

\makeatletter
\renewcommand{\@email}[1]{#1}
\makeatother

\emailAdd{\href{mailto:abissi@ictp.it}{\tt abissi@ictp.it},
            \href{mailto:fardelli@bu.edu}{\tt fardelli@bu.edu},
            \href{mailto:mkhansar@sissa.it}{\tt mkhansar@sissa.it}
}

\abstract{We investigate loop corrections to the four-point function of identical scalar operators in a four-dimensional large $N$ conformal field theory, holographically dual to AdS  with a quartic interaction.  We focus on the universal part of the correlator that, at any order in $1/N$, is completely determined by tree-level data. We show how this contribution controls the part of the anomalous dimensions of double-trace operators, that exhibits a characteristic 
 $\log \ell$ dependence at large spin $\ell$. We resum these effects to all orders in $1/N$  and  show that they admit a natural effective description in AdS.  Finally, by reformulating the problem in Mellin space, we demonstrate that the same contribution corresponds to consecutive unitarity cuts of bubble diagrams in the flat-space limit.
}

\begin{document}

\maketitle

\section{Introduction}
AdS/CFT is a duality mapping strongly coupled Conformal Field Theories (CFTs) to weakly coupled gravitational systems in asymptotically Anti-de Sitter (AdS) spacetimes, providing computational access to previously intractable non-perturbative regimes. Concurrently, the conformal bootstrap program leverages symmetry constraints, unitarity bounds, and crossing relations to delineate the landscape of consistent quantum field theories without recourse to traditional perturbative methods, see for instance \cite{Poland:2018epd,Bissi:2022mrs}.

The synergy between holographic techniques and bootstrap methods has yielded particularly significant insights into the structure of gravitational theories. In particular, there have been several advances in understanding how to study correlation functions in CFTs, with or without supersymmetry, in the holographic regime and as an expansion in large central charge, or large $N$. In particular, it is possible to use this approach to study holographic four-point functions. Starting with the seminal paper \cite{Heemskerk:2009pn}, there has been a lot of progress in understanding how to extend these results to higher order in the large $N$ expansion. An important methodological insight in this context has been the mapping of the singularities of the correlator in the regime in which the conformal cross ratio $v \to 0$, or equivalently $\bar{z} \to 1$, is compared to the CFT data \cite{Caron-Huot:2017vep,Alday:2015eya,Aharony:2016dwx}. Remarkably, in the first few orders in the large $N$ perturbation theory the singularities are controlled by the CFT data at the leading order, which makes it possible to completely reconstruct the correlator at order $N^{-4}$ \cite{Aharony:2016dwx,Fitzpatrick:2012cg}. 

The main obstacle to overcome to push this program further, is the appearance of higher-trace operators in the Operator Product Expansion (OPE), which start contributing to the singularity structure at order $N^{-4}$ and higher. However, at any given loop order in the large $N$ expansion, there is a term completely controlled by the CFT data of double-trace operators at leading orders. Although this term does not control the full singularity structure of the correlator, it provides interesting information about the full perturbation theory and on the respective holographic AdS loop amplitude. \\
In this paper, we study this term in the context of \cite{Heemskerk:2009pn} and focus on its effect to the anomalous dimension of double-trace operators.\footnote{In the case of $\mathcal{N}=4$ SYM, a similar setup has been studied in \cite{Bissi:2020woe,Bissi:2020wtv}.} We restrict ourselves to a specific kinematic limit, which corresponds to the large spin $\ell$ limit of the anomalous dimension. In particular, we note that at any given perturbative order $N^{-2\kappa}$ the anomalous dimension goes as $\frac{\log^{\kappa-2}\ell}{\ell^{2\Delta_{\CO}}}$ in the large spin limit, where $\Delta_{\CO}$ is the conformal dimension of the external operators in the four-point function. In this context, this behavior has not yet been observed, but it appeared for weakly coupled CFTs~\cite{Alday:2015ota,Gopakumar:2016wkt} and in the study of higher-trace operators~\cite{Fardelli:2025fkn,Henriksson:2023cnh}.\\
One of the objectives of this paper is to elucidate properties of the AdS amplitudes in connection to CFT features.  
This connection is made explicit through Witten diagrams, which provide the computational framework to evaluate AdS scattering amplitudes. These diagrams are the AdS analogue of Feynman diagrams in flat spacetime, but with crucial differences that reflect the geometry of AdS space and the holographic nature of the correspondence. The sum over all possible Witten diagrams for a given process reconstructs the complete CFT correlation function, providing an ecixplicit realization of the holographic principle in which bulk physics emerges from boundary correlations.

This correspondence reveals that the analytic structure of CFT correlation functions directly encodes the pole structure of AdS scattering amplitudes, which becomes even more transparent through the Mellin transform representation.
The relationship becomes particularly illuminating in the flat space limit, where AdS scattering amplitudes approach their Minkowski counterparts. In this limit, the conformal cross-ratios approach kinematic invariants of flat space scattering, and the conformal block expansion reproduces the partial wave expansion of scattering amplitudes. 

More specifically, we addressed the interplay between the contribution due to the most singular term of the four point function\footnote{This wording will become clear later, but with this we mean the term containing the highest power of $\log v$.} and the corresponding AdS loop amplitude.  We interpret the correction to the anomalous dimension of the double-trace operators as the correction due to the introduction of a new field in AdS at low twist, as expected from an effective field theory point of view. In particular, the all-loop resummation of the anomalous dimension on the CFT side exactly matches this interpretation. 

Another aspect that we analyzed is the Mellin respresentation of the most singular term. Similarly to \cite{Bissi:2020woe,Bissi:2020wtv}, we found an interesting interpretation in terms of consecutive unitarity cuts of the amplitude. We have shown this correspondence by inspecting the flat space limit of the Mellin amplitude and its genuine flat space computation. 

We list here some aspects that would be interesting to investigate in the future.
\begin{itemize}
\item The same analysis can be carried out in generic cases, for instance where there is no $\mathbb{Z}_2$ symmetry. The appearance of intermediate operators source another term in the double discontinuity which modifies the anomalous dimension of double-trace operators, see for instance \cite{Alday:2017gde}. It will be interesting to see how this modifies the large spin behavior of the anomalous dimension of double-trace operators and how to interpret it in the holographic context. 
\item It will be interesting to gain a better understanding of the behavior of higher-trace operators. Their study has recently attracted attention ~\cite{Fardelli:2025fkn,Fardelli:2025eun,Fardelli:2024heb,Kravchuk:2024wmv},providing new information about their spectrum. An interesting direction is to explore how their large-spin behavior relates to that of double-trace operators, for large enough twist and order in the large $N$ expansion,  and to clarify their interplay. A natural arena for such an analysis is provided by supersymmetric theories, where protected operators determine a larger fraction of the double discontinuity. Some studies along these lines are reported in \cite{Bissi:2024tqf,Bissi:2021hjk,Aprile:2024lwy,Aprile:2025hlt}. 
\end{itemize}

The paper is organized as follows.  In Sec.~\ref{sec1}  we begin with general definitions and notations, and provide a review of relevant results from the literature, which also serve as a testing ground to illustrate our methodology.  We finish the discussion of the CFT side with Sec.~\ref{sec2},  where we present our results for the corrections to the anomalous dimensions of double-trace operators at higher orders in the large $N$ expansion and the resummation of part of it.  Sec.~\ref{sec3} is devoted to the construction of the Mellin amplitude and to the interpretation of the CFT results obtained previously. The flat space limit and its interpretation in terms of cuts of the amplitude is presented in Sec.~\ref{sec4}. 

\section{Generalities}\label{sec1}
In this paper, we will focus on the study of the four-point function of  identical scalar operators with conformal dimension $\Delta_\CO$ in four dimensions $d$.  Conformal symmetry fixes its functional form to be
\eqna{
\langle \CO(x_1)\CO(x_2)\CO(x_3)\CO(x_4)\rangle=\frac{1}{(x_{12}^2 x_{34}^2)^{\Delta_\CO}}\CG(u,v)\, , 
}[4pt]
where we have defined $x_{ij}^2=(x_i-x_j)^2$ and $\CG$ is a function of conformal invariant cross-ratios
\eqna{
u=\frac{x_{12}^2x_{34}^2}{x_{13}^2x_{24}^2}\equiv z\zb\,,  \qquad\qquad v=\frac{x_{14}^2x_{23}^2}{x_{13}^2x_{24}^2}\equiv (1-z)(1-\zb)\, .
}[]
Invariance of the correlator under the exchange of any two operators translates in a set of crossing symmetry relations for $\CG$, which,  for $1\leftrightarrow 3$,  reads
\eqna{
\CG(u,v)=\left(\frac{u}{v} \right)^{\Delta_\CO} \CG(v, u)\, .
}[crossing]
Finally, $\CG$ admits an expansion in conformal blocks~\cite{Dolan:2003hv}
\eqna{
\CG(u,v)&=\sum_{\Delta, \ell}a_{\Delta, \ell} \lsp g_{\Delta, \ell} (u, v)\, ,\\
g_{\Delta, \ell} (u, v)&=\frac{z \zb}{z-\zb} \left( \kappa_{\Delta+\ell}(z)\kappa_{\Delta-\ell-2}(\zb)- z\leftrightarrow \zb \right)\, ,\\
\kappa_{\beta}(x)&=x^\frac{\beta}{2} {}_2F_1\left(\frac{\beta}{2}, \frac{\beta}{2}, \beta; x \right)\,,
}[blockExp]
where $\Delta$ and $\ell$ indicate the conformal dimension and spin of the operators exchanged in the OPE of $\CO \times \CO$ with OPE coefficient $f_{\CO\CO \CO_{\Delta, \ell}}$, such that  $a_{\Delta_\ell}=f_{\CO\CO \CO_{\Delta, \ell}}^2$.

To consider the simplest yet non-trivial setup, we study a large $N$ theory with a $\mathbb{Z}_2$ symmetry.  Holographically, this can be thought as a theory in AdS with only a quartic interaction, where the bulk coupling scales as $\sim \frac{1}{N^2}$.  Under these assumptions, the operators exchanged in the  $\CO\times \CO$ OPE are  
\eqna{
\CO\times \CO =\mathds{1}+[\CO\CO]_{n, \ell}+\cdots\, , 
}[OPE]
with $[\CO\CO]_{n, \ell}$ double-trace operators of schematic form
\eqna{
[\CO\CO]_{n, \ell}=\CO \lsp  \square\partial_{\mu_1} \cdots \partial_{\mu_\ell} \CO\, .
}[]
In fact, the $\mathbb{Z}_2$ symmetry forbids $\CO$, as well as any  multi-trace operators built from an  odd number of $\CO$,   from appearing in the $\cdots$ in~\eqref{OPE}.\footnote{Moreover, we can exclude the presence of the stress tensor $T$ by assuming that its self coupling with the scalars is much stronger than the gravitational coupling. This also removes from the $\cdots$ in \eqref{OPE},  any double-trace operators of the schematic type $[TT]$.} In the strict $N\to \infty$ limit,  only the identity and the double-trace operators appear in the OPE. When subleading $\frac{1}{N}$ corrections are included, these operators acquire corrections to their dimensions and OPE coefficients and additional multi-trace operators, with an even number of $\CO$ insertions,   can begin to appear in the OPE. 

To see how this structure emerges more concretely, we begin by considering the large $N$ expansion of the correlator 
\eqna{
\CG(u,v)=\CG^{(0)}(u,v)+\frac{1}{N^2} \CG^{(1)}(u,v)+\frac{1}{N^4} \CG^{(2)}(u,v)+\cdots\, , 
}[expCGLargeN]
where the leading term $\CG^{(0)}$ corresponds to Generalized Free Field  (GFF) theory. This can be  simply computed from Wick contractions and  matches the contributions of disconnected  AdS Witten diagrams
\eqna{
\CG^{(0)}=1+u^{\Delta_\CO}+\left(\frac{u}{v}\right)^{\Delta_{\CO}}\, .
}[]
By expanding this expression in conformal blocks according to~\eqref{blockExp}, we can directly see the existence of double-trace operators and read their free dimensions and OPE coeffiecients~\cite{Heemskerk:2009pn} 
\eqna{
\Delta_{n, \ell}^{(0)}&=2\Delta_{\CO}+2n+\ell\, ,  \\
a_{n, \ell}^{(0)}&=\frac{2(\ell +1)(2\Delta_\mathcal{O}+2n+\ell -2)}{(\Delta_\mathcal{O}-1)^2} C_n^{\Delta_\mathcal{O}-1} C_{n+\ell +1}^{\Delta_\mathcal{O}-1}\, , 
}[discOPE]
where by Bose symmetry only $\ell$ even are allowed and we have defined
\eqna{
C_n^{\Delta} = \frac{\Gamma^2(\Delta+n)\Gamma(2\Delta+n-1)}{\Gamma(n+1)\Gamma^2(\Delta) \Gamma(2\Delta+2n-1)}\,.
}[]
As anticipated,  at large $N$ these operators develop an anomalous dimension $\gamma$ and corrections to their three-point functions due to interactions
\eqna{
\Delta_{n, \ell}&=2\Delta_{\CO}+2n+\ell+\frac{\gamma_{n, \ell}^{(1)}}{N^2}+\frac{\gamma_{n, \ell}^{(2)}}{N^4}+\cdots\, , \\
a_{n, \ell}&=a_{n, \ell}^{(0)}+\frac{a_{n, \ell}^{(1)}}{N^2}+\frac{a_{n, \ell}^{(2)}}{N^4}+\cdots\, .
}[largeNDT]
Plugging this expansion in~\eqref{blockExp}, we can isolate for each $\CG^{(\kappa)}$ the contributions from double-trace operators. As an example we report the explicit expression for the first two orders
\twoseqn{
&\mathcal{G}^{(1)}=\sum_{n,\ell }u^{\Delta _{\mathcal{O} }+n}  \left(a^{(1)}_{n,\ell} +\frac{1}{2}a^{(0)}_{n,\ell} \gamma^{(1)}_{n,\ell} \left(\log (u)+\frac{\partial}{\partial n}\right)\right) \tilde{g}_{2 \Delta _{\mathcal{O} }+2 n+\ell,\ell}(u,v)\, ,
}[CG1]
{
&\begin{aligned}
\CG^{(2)}&=\sum_{n,\ell }u^{\Delta _{\mathcal{O} }+n}  \Bigg\lbrace \!  \left( a^{(2)}_{n,\ell } + \frac{1}{2} a^{(0)}_{n,\ell} \gamma^{(2)}_{n,\ell } \left( \log u + \frac{\partial}{\partial n} \right) \right)+ \frac{ a^{(1)}_{n\ell} \gamma^{(1)}_{n,\ell}}{2} \left( \log u + \frac{\partial}{\partial n} \right)
\\
&\quad\, +\frac{1}{8} a^{(0)}_{n,\ell} \left( \gamma^{(1)}_{n,\ell} \right)^2 \left( \log^2(u) + 2 \log u \frac{\partial}{\partial n} + \frac{\partial^2}{\partial n^2} \right)\! \Bigg\rbrace \tilde{g}_{2\Delta_\CO+2n+\ell, \ell}\, ,
\end{aligned}
}[CG2][GCBl]
where, to make manifest the leading small $u$ dependence,   we have redefined  the blocks
\eqna{
g_{\Delta, \ell}(z, \zb)=u^{\frac{\Delta-\ell}{2}}\tilde{g}_{\Delta, \ell}(z, \zb)\, .
}[]
If we specify to an AdS theory with only a quartic interaction, it is  known~\cite{Heemskerk:2009pn} that at order in $N^{-2}$ only double-trace operators with $\ell=0$ get corrections.\footnote{This result can also be explicitly verified by expanding in blocks $\CG^{(1)}$ given by  the corresponding  Witten diagram,  ie  $\CG^{(1)}\sim \Db_{\Delta_\CO\Delta_\CO\Delta_\CO\Delta_\CO}$.} Up to an overall coefficient,\footnote{We fixed the overall coefficient by demanding that $\gamma^{(1)}_{0,0}=1$. This can be achieved by a re-parametrization of $N$.} these corrections are
\twoseqn{
 \gamma^{(1)}_{n,0}&=\frac{(2 \Delta_\mathcal{O}-1) (n+1) (2 \Delta_\mathcal{O}+n-3) (\Delta_\mathcal{O}+n-1)}{(\Delta_\mathcal{O}-1) (2 \Delta_\mathcal{O}+2 n-3) (2 \Delta_\mathcal{O}+2 n-1)}\, ,
}[treeOPE]
{
  a^{(1)}_{n,0}&=\frac{1}{2} \partial_n(  a^{(0)}_{n,0} \gamma^{(1)}_{n,0} )\, .
}[][]

So far, we have relied on the explicit expansion in conformal blocks to extract operator dimensions and OPE coefficients. However, there exists a powerful alternative technique --- eventually the only viable approach for accessing higher-order corrections in the $1/N$ expansion --- which is the Lorentzian inversion formula~\cite{Caron-Huot:2017vep} (LIF).  The power of the LIF lies in the fact that it encodes OPE data in a function $c(\Delta, \ell)$, known as OPE coefficient density,  which depends not on the full correlator, but only on its singular behavior as $\zb\to 1$. Technically, this is realized by expressing  $c(\Delta, \ell)$ not in terms of $\CG(z, \zb)$, but rather in terms of its double discontinuity (dDisc), defined as the difference between the Euclidean correlator and its two analytic continuations around $\zb=1$
\eqna{
\dDisc\left[ \CG(z, \zb)\right]=\CG(z, \zb)-\frac{1}{2}\CG^\circlearrowright(z, \zb)-\frac{1}{2}\CG^\circlearrowleft(z, \zb)\, .
}[]
For the correlator of identical operators in $4d$, the OPE coefficient density is
\eqna{
c(\Delta, \ell)=\frac{1+(-1)^\ell}{4}k_{\frac{\Delta+\ell}{2}} \int_{0}^1 dz d\zb \frac{(z-\zb)^2}{z^4\zb^4}g_{\ell+3, \Delta-3}(z, \zb)\dDisc\left[ \CG(z, \zb)\right]\, ,
}[LIFc]
with $k_\alpha=\frac{\Gamma^4(\alpha)}{2\pi^2 \Gamma(2\alpha-1)\Gamma(2\alpha)}$.  Crucially, the dimensions and OPE coefficients of the exchanged operators are then encoded, respectively, in the poles and residues of  of $c(\Delta, \ell)$
\eqna{
\lim_{\Delta\to\Delta_*}c(\Delta, \ell)=\frac{a_{\Delta_*, \ell}}{\Delta-\Delta_*}\, .
}[cResOPE]
The function $c(\Delta, \ell)$ inherits from  $\CG$ in~\eqref{expCGLargeN} a large $N$ expansion, such that~\eqref{cResOPE} can be expressed, order by order,  in terms of corrections to the OPE data in~\eqref{largeNDT}
\eqna{
c(\Delta, \ell)&=\frac{a^{(0)}_{n, \ell}}{\tau-\tau_{n}^{(0)}}+\frac{1}{N^2}\left( \frac{a^{(0)}_{n, \ell}\gamma^{(1)}_{n, \ell}}{(\tau-\tau_{n}^{(0)})^2}+\frac{a^{(1)}_{n, \ell}}{\tau-\tau_{n}^{(0)}}\right) +\frac{1}{N^4}\Bigg( \frac{a^{(0)}_{n, \ell}(\gamma^{(1)}_{n, \ell})^2}{(\tau-\tau_{n}^{(0)})^3}\\
&\quad\, +\frac{a^{(0)}_{n, \ell}\gamma^{(2)}_{n, \ell}+a^{(1)}_{n, \ell}\gamma^{(1)}_{n, \ell}}{(\tau-\tau_{n}^{(0)})^2}+\frac{a^{(2)}_{n, \ell}}{\tau-\tau_{n}^{(0)}}\Bigg)+\cdots
}[cExp]
where we have expressed everything in terms of the twist $\tau=\Delta-\ell$ and $\tau_n=2\Delta_{\CO}+2n$.\footnote{In our example, all the terms containing $\gamma_{n,\ell}^{(1)}$ and $a_{n,\ell}^{(1)}$  being non zero only for $\ell=0$ do not  produce any pole. }

To apply the LIF we will first rewrite the correlator in terms of its crossed-channel version. In other words, instead of using  $\CG$ directly in~\eqref{LIFc}, we  substitute 
\eqna{
\CG(z, \zb)=\left(\frac{z\zb}{(1-z)(1-\zb)}\right)^{\Delta_\CO} \CG(1-z, 1-\zb).
}[]
With this transformation, the only relevant terms with non-vanishing dDisc are 
\eqna{
\dDisc\left[\left(\frac{1-\zb}{\zb} \right)^\lambda\right]&=\left(\frac{1-\zb}{\zb} \right)^\lambda 2\sin(\pi\lambda)^2\, , \\
\dDisc\left[ \log^n(1-\zb)\right]&=2\pi^2 n(n-1)\log^{n-2} (1-\zb)+\text{lower powers of }\log(1-\zb)\, , 
}[]
where $\lambda<0$ and  notice that for a single $\log(1-\zb)$ dDisc is zero.  
Given these results, it is straightforward to verify that the disconnected OPE data in~\eqref{discOPE} are reproduced by considering only the constant term in $\CG^{(0)}$.  Upon crossing, this term becomes $\left(\frac{u}{v}\right)^{\Delta_\CO}$, which is the only singular contribution and thus the only one with a non-vanishing double discontinuity.  Moving on to the $\CO(N^{-2})$ correction,   we need to analyze the singular behavior of $\CG^{(1)}(v, u)$ defined  in~\eqref{CG1}. To do so, it is important to understand the behavior of each conformal block as $v\to 0$
\eqna{
\tilde{g}_{2\Delta_{\CO}+2n+\ell, \ell}(u,v) \stackrel{v\to 0}{\approx}f(u,v)+\tilde{f}(u,v)\log v\, ,
}[]
where both $f(u,v)$ and $\tilde{f}(u,v)$ admit a Taylor expansion around small $u$ and $v$.  Thus, each individual block diverges at most logarithmically. However, summing over an infinite tower of spins can potentially change this behavior.  At order $1/N^2$,  only double-trace operators with spin zero acquire anomalous dimensions, so no logarithmic enhancement from the large-spin tail is expected. As a result, we  expect 
\eqna{
\left(\frac{u}{v}\right)^{\Delta_\CO }\CG^{(1)}(v, u) \approx v^n \left( f_1(u) \log u+ f_2(u) \log v +f_3(u) \log u \log v+f_4(u)\right)\, .
}[]
Since  $\dDisc[\log v]=0$,  it may seem that no singular terms survive and the Lorentzian inversion formula fails to reproduce the correct 
 $1/N^2$ OPE data in~\eqref{treeOPE}. This apparent contradiction is resolved by noting that the inversion formula is only guaranteed to reproduce the correct OPE data for non truncated in spin $\gamma_{n,\ell}$. In our case,  $\gamma_{n, \ell}^{(1)}$ and $a_{n, \ell}^{(1)}$ are nonzero only for $\ell=0$.  For similar reasons, at the next order, the only term in $\CG^{(2)}(v,u)$ surviving the  double-discontinuity  is
\eqna{
\frac{1}{8}u^{\Delta_\CO}\log^2 v \sum_{n}a_{n, 0}^{(0)} (\gamma_{n,0}^{(1)})^2v^n\tilde{g}_{2\Delta_\CO+2n, 0}(v,u)\, .
}[logv2]
Notice that this term is fully determined by the OPE data at the previous order and, quite remarkably --- as we will soon see --- it is sufficient to completely fix $\gamma_{n, \ell}^{(2)}$ and $a_{n, \ell}^{(2)}$.

In the remainder of this section, we will carefully go through the procedure to extract the OPE data at order $N^{-4}$ from $c(\Delta, \ell)$, as in~\eqref{cExp}, by plugging~\eqref{logv2} into~\eqref{LIFc} and evaluating the resulting inversion integral. We will then show that this reproduces the results previously obtained in~\cite{Aharony:2016dwx}.   Since here and in the following, we will focus on the lowest-twist double-trace operators, ie  at $n=0$, we can simplify~\eqref{LIFc} by expanding in collinear blocks~\cite{Caron-Huot:2017vep,Alday:2017zzv,Henriksson:2020jwk} and work with
\eqna{
\tilde{c}(h, \hb)=\int_{0}^1 \frac{dz}{z}z^{-h} \int_{0}^1\frac{d\zb}{\zb^2}  k_{2\hb}\lsp  \kappa_{2\hb}(\zb) \dDisc\left[ \CG(z, \zb)\right]_{z^{h_0}}
}[]
where we have defined $h=\frac{\tau}{2}$ and $\hb=\frac{\tau+2\ell}{2}$ and by $[\, ]_{z^{h_0}}$ we denote the projection onto the term of order $z^{h_0}$  in the Taylor expansion around $z= 0$.  At order $N^{-4}$, the only term with non-vanishing dDisc is~\eqref{logv2} and if we set $\Delta_{\CO}=2$ and select the $z^2$ term we get
\eqna{
&\dDisc\left[\frac{1}{8}u^2 \log^2 v \sum_n a_{n,0}^{(0)} (\gamma_{n,0}^{(1)})^2\tilde{g}_{4+2n, 0}(v,u)\right]_{z^2}=s(\zb)\log z +t(\zb) \, ,\\
&s(\zb)=\sum_{m=0}^\infty s_m \zb (1-\zb)^m  \, , \quad s_m=\begin{cases}
-6\pi^2 & \,m=0\, ,\\
\sum\limits_{j=0}^m \frac{-36 \pi ^2 (j+1)^8 \Gamma (m+1)^2}{(2 j+1) (2 j+3) \Gamma (-j+m+1) \Gamma (j+m+3)} & \text{ else}\, .
\end{cases}
}[]
Plugging this expression into the inversion integral, we can first focus on the $z$-integral
\eqna{
\int \frac{dz}{z }z^{-h} z^{h_0}\log^p z =-\frac{\Gamma(p+1)}{(h-h_0)^{p+1}}\, .
}[]
Comparing with~\eqref{cExp}, we see that extracting  $a_{0,\ell}^{(0)}\gamma_{0, \ell}^{(2)}$  requires considering only the $\log z$ part, as it is the only term that produces a double pole in the inversion formula.  In formulae, this means we can write
\eqna{
\gamma_{0, \ell}^{(2)}=-4 \frac{1}{a_{0, \ell}^{(0)}}\int\frac{d\zb}{\zb^2}  k_{2\hb}\lsp  \kappa_{2\hb}(\zb)s(\zb)
}[]
where the $4$ comes from the fact that $h=\frac{\tau}{2}$. To make manifest the dependence of $\hb$ or equivalently of the spin $\ell$ it is convenient to rewrite $s(\zb)$ as an expansion of powers of $\frac{1-\zb}{\zb}$
\eqna{
s(\zb)=\sum_{k=0}^\infty \tilde{s}_k \left( \frac{1-\zb}{\zb} \right)^k \, ,
}[]
where $\tilde{s}_k$ are simply related to $s_m$, for instance
\eqna{
\tilde{s}_0=s_0\,, \qquad \tilde{s}_0=-s_0+s_1\, , \qquad \tilde{s}_2=s_0-2s_1+s_2 \,  .
}[]
In this way, the integrals become trivial and we obtain
\eqna{
\gamma_{0, \ell}^{(2)}&=\sum_{k}\tilde{s}_k \frac{ \Gamma (k+1)^2 \Gamma (\hb-k-1)}{\pi ^2 (\hb-1) \hb \Gamma (\hb+k+1)}\, .
}[gammaw]
An advantage of this representation is that it makes the large spin behavior manifest.  Indeed, each term in the sum behaves at large  ``conformal spin'' $J^2=\hb(\hb-1)=(\ell+\Delta_{\CO}+n)(\ell+\Delta_{\CO}+n-1)$, ie. $J^2=(\ell+2)(\ell+1)$ for the lowest twist,  as $J^{-4-2k}$ for $J\gg 1$. More explicitly
\eqna{
\gamma_{0, \ell}^{(2)}\xrightarrow[]{J\to\infty}-\frac{6}{J^4}\left(  1+\frac{18}{5}\frac{1}{J^2}+\frac{96}{7}\frac{1}{J^4}+\frac{360}{7}\frac{1}{J^6}+\frac{74304}{385}\frac{1}{J^8}\right)+\cdots
}[gamma2] 
in accordance with~\cite{Aharony:2016dwx}.  Similarly we can extract the correction to the OPE coefficient integrating $t(\zb)$
\eqna{
\frac{a^{(2)}_{0, \ell}}{a_{0,\ell}^{(0)}}\xrightarrow[]{J\to\infty}\frac{6}{J^4}\left(  1 +\frac{59}{10}\frac{1}{J^2}+\frac{1048}{35}\frac{1}{J^4}\right)+\cdots
}[a2]
\section{Higher orders}\label{sec2}
\subsection{Anomalous dimensions}
At order $1/N^2$, the contribution in~\eqref{logv2} captures the full double discontinuity of the correlator and therefore determines both
$\gamma_{0, \ell}^{(2)}$ and $a^{(2)}_{0, \ell}$ completely.  At higher orders this is no longer the case: dDisc receives contributions not only from tree-level data,  and in general fixing the corrections at order 
 $1/N^{2\kappa}$ requires knowledge of the full set of OPE data at order $1/N^{2(\kappa-1)}$.  Nevertheless, there is always a universal part of the correlator that depends only on the leading OPE coefficients  $a^{(0)}_{n,0}$ and tree-level anomalous dimensions $\gamma^{(1)}_{n,0}$, which produces a  non-vanishing dDisc. This contribution is the higher-order analogue of~\eqref{logv2}, namely
\eqna{
\CG^{(\kappa)} \supset \frac{1}{2^\kappa \kappa!}u^2 \log^\kappa v  \sum_{n}a_{n, 0}^{(0)} (\gamma_{n,0}^{(1)})^{\kappa}v^n\tilde{g}_{4+2n, 0}(v,u)\, ,
}[leadinglogs]
where we have set $\Delta_{\CO}=2$.  While this does not reproduce the full dDisc, we will see that these ``leading-log'' contributions still control a specific part of the OPE data.  To distinguish the data extracted by inverting only this partial contribution from the full result, we will denote them with a $\hat{\cdot}$.\\
To see how this mechanism works, let us first project onto the leading-twist sector.  Then,  similarly to the  $\kappa=2$ case, we find 
\eqna{
&\text{dDisc}\left[\frac{1}{2^\kappa \kappa!}u^2 \log^\kappa v  \sum_{n}a_{n, 0}^{(0)} (\gamma_{n,0}^{(1)})^{\kappa}v^n\tilde{g}_{4+2n, 0}(v,u)\right]_{z^2}&\\
&\qquad\qquad=2\pi^2 \kappa(\kappa-1)\log^{\kappa-2}(1-\zb)\left( s^{(\kappa)}(\zb)\log z+t^{(\kappa)}(\zb)\right)+\cdots \, .
}[dDiscLeadingLogs]
The appearance of the $\log z $ term produces double poles, which can be directly identified with the contributions to $\tilde{c}^{(\kappa)}(h, \hb)$ of the schematic form
\eqna{
\frac{1}{(\tau-4)^2}\left(a^{(0)}_{0, \ell} \hat{\gamma}^{(\kappa)}_{0, \ell} +\sum_{i=2}^{\kappa-2}\hat{a}_{0, \ell}^{(i)}\hat{\gamma}_{0, \ell}^{(\kappa-i)}\right)\, ,
}[doublepole]
from which we can extract  $\hat{\gamma}^{(\kappa)}$.  Proceeding order by order, we find 
\eqna{
\hat{\gamma}_{0, \ell}^{(3)}\xrightarrow[]{J\to\infty}\frac{6}{J^4}\left(\log J+\gamma_E \right)+\frac{1}{25J^6}\left( 924(\log J+\gamma_E)-899\right)+\cdots\, ,
}[]
where $\gamma_E$ is the Euler-Mascheroni constant.  In addition,  for $\kappa=3$, the analysis of the simple pole yields the correction to the OPE coefficient as well
\eqna{
\frac{\hat{a}^{(3)}_{0, \ell}}{a^{(0)}_{n, \ell}}\xrightarrow[]{J\to\infty}-\frac{6}{J^4}(\log J+\gamma_E)+\cdots\, .
}[]
A remarkable feature of these results is the  appearance  of  $\log J$,  unlike the results at previous order. This is significant for two reasons.  First,  from the inversion formula, terms of the form $\log^{\kappa-2} J$ can only arise from the integration of $\log^{\kappa-2}(1-\zb)$,  which are completely encoded in~\eqref{dDiscLeadingLogs}. It follows that the $\log J$ part of $\hat{\gamma}^{(3)}$ exactly matches the corresponding piece of the full anomalous dimensions, and more generally 
\eqna{
\hat{\gamma}^{(\kappa)}_{0, \ell}\Big|_{\log^{\kappa-2}J}={\gamma}^{(\kappa)}_{0, \ell}\Big|_{\log^{\kappa-2}J}
}[]
with an analogous statement for the OPE coefficients.
Second, this behavior could have been anticipated from the structure of the correlator and in particular its behavior under crossing symmetry. In fact,  when expanded in conformal blocks,  $\CG^{(3)}(u,v)$ contains a contribution of the form 
\eqna{
\CG^{(3)}(u,v)\supset \frac{1}{2}u^2 \log u \sum_{\ell}a^{(0)}_{0, \ell}\gamma^{(3)}_{0, \ell} \tilde{g}_{4+\ell, \ell}(u, v)\, .
}[]
At large $\ell$, or rather large $J$, this contribution leads to~\cite{Alday:2016njk} 
\eqna{
 \log u  \sum_{J}a^{(0)}_{0, J} \frac{\log J}{J^2} \tilde{g}_{4+J, J}(u, v)\approx  \log u \log^3 v\, .
}[]
This $\log v$ originates exclusively from the $\log J$ term and, upon crossing,  provides precisely the contribution required to reproduce the $a^{(0)}(\gamma^{(1)})^3 \log^3 u \log v$ structure in~\eqref{leadinglogs}.\footnote{A similar $\frac{\log J}{J^2}$ behavior can be observed  in the tree-level anomalous dimensions of quadrupole-trace operators due to a $\phi^4$ interaction.  These operators indeed start to enter the OPE at $1/N^4$ and mix with double-trace operators starting from twist 8 for $\Delta_{\CO}=2$.} For the same reason at higher order we expect $\log^{\kappa-2} J$.   Starting from $\kappa=4$, however, the situation becomes more subtle because the residue of the double pole contains not only the new correction to the anomalous dimension but also contributions proportional to lower-order data. Concretely, at 
 $\kappa=4$,  as follows from~\eqref{doublepole}, the relevant structure is
\eqna{
\hat{\gamma}^{(4)}_{0, \ell}+\frac{\hat{a}_{0, \ell}^{(2)}\hat{\gamma}^{(2)}_{0, \ell}}{{a}_{0, \ell}^{(2)}}\, .
}[logJgen]
Nevertheless, using the explicit results of~\eqref{gamma2} and~\eqref{a2}, one sees that the second term cannot generate
 $\log^2 J$. Hence any  $\log^2 J$ produced by the inversion formula  at this order must come entirely from $\hat{\gamma}^{(4)}_{0, \ell}$.  And by the same logic as above,  this contribution coincides with the corresponding piece of the full anomalous dimension.  In general, we can therefore state that the residue of the double pole coming from~\eqref{dDiscLeadingLogs} provides the leading large-$J$ behavior  of $\gamma^{(\kappa)}_{0, \ell}$
\eqna{
 \gamma^{(\kappa)}_{0, \ell}\approx \log^{\kappa-2}\!J \left(\frac{\delta^{(\kappa)}_1}{J^4}+\frac{\delta^{(\kappa)}_2}{J^6}+\frac{\delta^{(\kappa)}_3}{J^8}+\cdots\right)+\cdots\, ,
}[]
where explicit evaluation for several values of $\kappa$ yields
\eqna{
\delta^{(\kappa)}_1&=\frac{6(-1)^{\kappa+1}}{\Gamma(\kappa-1)}\, , \\
\delta^{(\kappa)}_2&=\frac{4(-1)^{\kappa}}{\Gamma(\kappa-1)}\left(1-4 \left( \frac{8}{5}\right)^{\kappa-1} \right)\, , \\
\delta^{(\kappa)}_3&=\frac{4(-1)^{\kappa+1}}{\Gamma(\kappa-1)}\left(1-7 \left( \frac{8}{5}\right)^{\kappa}+7 \left( \frac{81}{35}\right)^{\kappa} \right)\, .
}[]
Reinstating the factors of $1/N$, this series can actually be resummed  in $\kappa$, giving the leading large $J$ contribution to the leading twist anomalous dimension $\gamma_{0, \ell}\equiv \gamma_{\ell}$
\eqna{
\gamma_{\ell}(N)&\approx \frac{1}{N^4}\Bigg[- \frac{6J^{-\frac{1}{N^2}}}{J^{4}}+\frac{4}{J^6}\left(J^{-\frac{1}{{N}^2}}-\frac{32}{5} J^{-\frac{8}{5 {N}^2}} \right)\\
&\quad\, -\frac{4}{J^8}\left( J^{-\frac{1}{N^2}}-\frac{448}{25} J^{-\frac{8}{5 N^2}}+\frac{6561}{175} J^{-\frac{81}{35 N^2}}\right)\Bigg]+\cdots\, .
}[]
Here the powers of $J$ can be recognized as precisely the tree level  anomalous dimensions $\gamma_{n, 0}^{(1)}$ defined in~\eqref{treeOPE},  
\eqna{
\gamma_{0, 0}^{(1)}=1\, , \qquad \quad \gamma_{1, 0}^{(1)}=\frac{8}{5}\, , \qquad \quad \gamma_{2, 0}^{(1)}=\frac{81}{35}\, .
}[]
This observation leads us to a compact general formula
\eqna{
\gamma_{\ell}(N)&\approx\frac{1}{N^4} \sum_{n, m} \frac{f_{nm}}{J^{4+2n+2m+\frac{\gamma_{n,0}^{(1)}}{N^2}}}\, ,
}[gammaN]
with coefficients $f_{nm}$ independent of $N$.  In the next section, we will  interpret this result from an effective field theory point of view. 
\subsection{Interpretation as a scalar exchange in AdS}
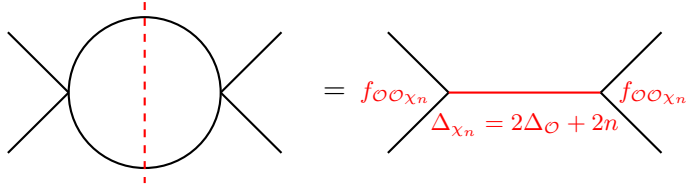
\begin{figure}
\centering
\begin{tikzpicture}
\draw[thick] (0,0) circle (1);
\draw[thick] (-1,0) -- (-1.8,0.8);
\draw[thick] (-1,0) -- (-1.8,-0.8);
\draw[thick] (1,0) -- (1.8,0.8);
\draw[thick] (1,0) -- (1.8,-0.8);
\draw[thick, red, dashed] (0,1.2) -- (0,-1.2);
\node at (2.5, 0) {$=$};
\draw[thick] (4,0) -- (3.2,0.8);
\draw[thick] (4,0) -- (3.2,-0.8);
\draw[thick,red] (4,0) -- (6,0);
\draw[thick] (6,0) -- (6.8,0.8);
\draw[thick] (6,0) -- (6.8,-0.8);
\node[below=0.15cm, red] at (5,0) {\footnotesize $\Delta_{\chi_n}=2\Delta_{\CO}+2n$};
\node[right=0.1cm, red] at (6,0) {\footnotesize $f_{\CO\CO{\chi_n}}$};
\node[left=0.1cm, red] at (4,0) {\footnotesize $f_{\CO\CO{\chi_n}}$};
\end{tikzpicture}
\caption{Spectral decomposition of bubble diagrams as a sum over tree-level scalar exchanges. }\label{Fig: SD}
\end{figure}
To understand how an effective description emerges, let us return to the case $\kappa=2$.  In this situation, the one-loop bubble diagram can be decomposed, via the spectral representation of~\cite{Fitzpatrick:2011dm}, into a sum over tree-level scalar exchanges ${\chi_n}$, with dimension $2\Delta_\CO+2n$,  as illustrated in Fig.~\ref{Fig: SD}.  At this order,   using the appropriate OPE coefficients and setting $\Delta_{\chi_n}=4+2n$,  one recovers the result for $\gamma^{(2)}_{0, \ell}$  in~\eqref{gamma2},  where the exchange of the scalar of dimension $4+2n$   contributes  starting from $1/J^{4+2n}$. At higher order in $1/N$,  we would like to push this analogy further and formulate an effective field theory at large spin. The idea is to not consider  explicit loop computations and instead promote each of the scalars $\chi_n$ to dynamical fields in AdS with shifted dimensions $\Delta_{\chi_n}=4+2n+\frac{1}{N^2}\gamma_{n,0}^{(1)}$.  At large spin, the anomalous dimension of leading-twist double-trace operators induced by the exchange of an AdS scalar takes the form~\cite{Fitzpatrick:2012yx}
\eqna{
\gamma_{\ell} \approx -\frac{1}{J^{\Delta_{\chi_n}}}\frac{2f_{\CO\CO\chi_n}^2 \Gamma(\Delta_\CO)^2\Gamma(\Delta_{\chi_n})}{\Gamma\mleft(\Delta_\CO-\frac{\Delta_{\chi_n}}{2}\mright)^2\Gamma\mleft(\frac{\Delta_{\chi_n}}{2}\mright)^2}\, .
}[largeJspin]
Focusing for simplicity on the case $n=0$, and substituting the large-$N$ expansion of the OPE coefficient and the shifted dimension, one immediately reproduces the leading  $\log J$ term multiplying $1/J^4$  
\eqna{
\gamma_{\ell}\approx- \frac{3}{J^{4}}\sum_{\kappa=2}\frac{1}{N^{2\kappa}} a^{(0)}_{0,0} \left( \gamma_{0,0}^{(1)}\right)^{\kappa} \frac{(-1)^{\kappa}\log^{\kappa-2} J}{\Gamma(\kappa-1)}\, ,
}[largeJscalar]
where we have used  $\gamma_{0,0}^{(1)}=1$ and $a^{(0)}_{0,0}=2$.  
Similar logic applies for $n \geq 1$. In this case, the expression in~\eqref{largeJspin} does not fully determine the coefficient of the $1/J^{4+2n}$ term, but it does fix the universal contribution corresponding to
\eqna{
-\frac{1}{N^4}\frac{36 (n+1)^8 \Gamma (n+1)^4}{(2 n+1) \Gamma (2 n+4)}\frac{1}{J^{4+2n+{\gamma_{n,0}^{(1)}}/{N^2}}}\,.
}[]
\subsection{Subleading twist}
So far our analysis has focused on the leading-twist operators, but the same procedure can be applied to double-trace operators with higher twists. As an illustration, consider the case $n=1$.  By extracting the $z^3$ contribution from dDisc,   one finds at large spin, for the first few orders in $1/N$,
\eqna{
\gamma^{(2)}_{1, \ell}&\approx -\frac{48}{J^4}\left( 1+\frac{51}{5J^2}+\cdots \right)\, ,\\
\hat{\gamma}^{(3)}_{1, \ell}&\approx  \frac{48}{J^4}\log J \left( 1+\frac{383}{25J^2}+\cdots \right)+\cdots\, , \\
\hat{\gamma}^{(4)}_{1, \ell}&\approx - \frac{8}{J^4}\log^2 J \left( 3+\frac{8817}{125J^2}+\cdots \right)+\cdots\, , \\
}[]
where we have omitted terms suppressed by lower powers of $\log J$ and in this case $J^2= (\ell+3)(\ell+2)$. 

It is important to mention that starting from twist eight,  the anomalous dimensions extracted from the double poles of the inversion formula may not coincide directly with those of the pure double-trace operators.  In fact, at $1/N^4$ order and beyond,  quadrupole and higher-trace operators also appear in the OPE and mix with double traces of the same classical dimensions.\footnote{Recall that we define quadrupole-trace operators and in general $Q$-trace operators as operators with $\tau=Q \Delta_\CO+2n$.} As a result, what is obtained from the double-pole structure is an average anomalous dimension over all such operators, and a proper diagonalization of the mixing problem would be required to isolate the individual contributions.
\section{Mellin space}\label{sec3}
An alternative and particularly powerful representation of correlation functions is provided by Mellin space.  In this formulation, correlation functions acquire a simple analytic structure in terms of Mellin variables, which supports their interpretation as scattering amplitudes in AdS and  closely parallels that of flat-space amplitudes, to which they reduce in the appropriate limits~\cite{Penedones:2010ue,Mack:2009mi,Mack:2009gy,Fitzpatrick:2011ia,Paulos:2011ie,Nandan:2011wc}.

Given the definition of the four-point function in~\eqref{4pt}, the corresponding Mellin amplitude is defined as~\cite{Penedones:2010ue}
\eqna{
\CG(u, v)=\!\!\int_{-i \infty}^{i\infty}\frac{ds dt}{(4\pi i)^2} \lsp u^{\frac{t}{2}}v^{\frac{\hat{u}-2\Delta_{\CO}}{2}} \Gamma\mleft(\frac{2\Delta_{\CO}-t}{2}\mright)^2\! \Gamma\mleft(\frac{2\Delta_{\CO}-s}{2}\mright)^2\! \Gamma\mleft(\frac{2\Delta_{\CO}-\hat{u}}{2}\mright)^2 \!M(s, t)\, ,
}[mellin]
where $s, \, t, \, \hat{u}$ are Mandelstam-like variables satisfying $s+t+\hat{u}=4\Delta_{\CO}$.  In Mellin space, the crossing-symmetry relations  take the form
\eqna{
M(s,t)=M(s, \hat{u})=M(t,s)\, .
}[]
Finally the Mellin amplitude $M(s,t)$ inherits the large $N$ expansion of $\CG(u,v)$ given in~\eqref{expCGLargeN}
\eqna{
M(s, t)=\frac{M_{\rm tree(s, t)}}{N^2}+\frac{M_{\rm 1-loop}(s, t)}{N^2}+\cdots\, ,
}[]
where the subscript ``tree'' denotes the Mellin transform of $\CG^{(1)}$, and more generally we denote by $M_{(\kappa-1)\text{-loop}}$  the Mellin amplitude of $\CG^{(\kappa)}$.  While it is well known that the tree-level Mellin amplitude associated with an AdS $\phi^4$ 
  interaction is simply a constant, our aim here is to construct the higher-loop contributions in such a way that they reproduce the expected double-trace expansion in~\eqref{CG2}.  For example, at one-loop,  reproducing the $\log^2u$ term requires a Mellin amplitude of the form\footnote{Any regular contribution of the type $f_{\rm}(s,t)$
 can in principle be added to this ansatz, the effect being equivalent to a choice of renormalization scheme. Throughout this work we fix this ambiguity by setting $f_{\rm}(s,t)=0$.}
\eqna{
M_{\rm 1-loop}=\sum_{m=0}^\infty \frac{R_m^{(1)}(s)}{t-(2\Delta_{\CO}+2m)}+(s\leftrightarrow t)+(t\leftrightarrow \hat{u}) \, .
}[]
Substituting this ansatz into~\eqref{mellin}, it is straightforward to check that it leads to the expected structure
\eqna{
\underset{t=2\Delta_{\CO}+2m}{\text{Res}}\left[ \CG^{(2)}(u,v)\right]=u^{\Delta_{\CO}+m}\left( A_m(v)\log^2u+B_m(v)\log u+C_m(v)\right)\, ,
}[]
where the lhs denotes the residues when $\CG^{(2)}(u,v)$ is written in terms of its Mellin amplitude.
In~\cite{Aharony:2016dwx} it was shown how to determine the coefficients $R_m^{(1)}$ in this particular case and that an ansatz of this type completely fixes the Mellin amplitude at this order.  For higher loops,  we will instead focus on constructing a Mellin amplitude that reproduces only the leading-log contribution in~\eqref{leadinglogs}, which takes the form
\eqna{
M^R_{(\kappa-1)\text{-loop}}(s,t)=\sum_{m=0}^\infty \frac{R_m^{(\kappa-1)}(s)}{(t-(2\Delta_{\CO}+2m))^{\kappa-1}}+\text{crossed}\, .
}[MR]
Although this is only part of the full solution, it already captures significant physical information. In what follows, we show how the coefficients $R_m^{(\kappa-1)}$  can be derived from tree-level data and how they reproduce the anomalous dimensions $\gamma^{(\kappa)}_{0, \ell}$ obtained previously.
\subsection{Extracting the $R_m^{(\kappa-1)}$ residues}
To determine the residues in~\eqref{MR}, it is useful to recall the analogue of the conformal block decomposition for the Mellin amplitude.  In the $t$-channel, this takes the form
\eqna{
M(s,t)=\sum_{p}a_{\Delta_p, \ell_p}\sum_{m=0}^\infty \frac{\CQ_{\ell_p,m}(s, \tau_p)}{t-(\tau_p+2m)}\, ,
}[]
where the sum over $p$ runs over primary operators $\CO_p$, with dimension $\Delta_p$ and spin $\ell_p$,  exchanged in the $\CO\times\CO$ OPE, while the sum over $m$ accounts for their descendants. The functions $\CQ_{\ell_p,m}(s, \tau_p)$ are known as Mack polynomial~\cite{Mack:2009gy}, which can be written as~\cite{Costa:2012cb} 
\eqna{
\CQ_{\ell_p, m}(s, \tau_p)=\frac{-2\Gamma(\Delta_p+\ell_p)(\Delta_p-1)_{\ell_p}}{4^{\ell_p} m! (\Delta_p-1)_m \Gamma\mleft(\frac{\Delta_p+\ell_p}{2}\mright)^4\Gamma\mleft(\Delta_{\CO}-m-\frac{\Delta_p-\ell_p}{2}\mright)^2}Q_{\ell_p, m}(s, \tau_p) \, ,
}[defQ]
where $Q_{\ell_p, m}(s, \tau_p)$ is a polynomial of degree $\ell_p$ in $s$ with $Q_{0, m}(s, \tau_p)=1$. 

Specializing to the case where $\CO_p$ are the double-trace operators in~\eqref{largeNDT}
\eqna{
M(s,t)=\sum_{m,n}  \frac{\left( a_{n, \ell}^{(0)}+\frac{a_{n, \ell}^{(1)}}{N^2}+\cdots\right)\CQ_{\ell, m}\left(s, 2\Delta_{\CO}+2n+\frac{\gamma_{n, \ell}^{(1)}}{N^2}+\frac{\gamma_{n, \ell}^{(2)}}{N^4}+\cdots\right)}{t-\left(2\Delta_{\CO}+2n+2m+\frac{\gamma_{n, \ell}^{(2)}}{N^2}+\frac{\gamma_{n, \ell}^{(1)}}{N^4}+\cdots \right)}+\cdots\,.
}[Mexp]
We first expand the Mack polynomial and by using~\eqref{defQ} it is straightforward to verify that
\eqna{
\CQ_{\ell, m}\Big(s, 2\Delta_{\CO}+2n+\frac{\gamma_{n, \ell}^{(1)}}{N^2}+\frac{\gamma_{n, \ell}^{(2)}}{N^4}+\cdots\Big) \propto \left(\frac{\gamma_{n, \ell}^{(1)}}{N^2}\right)^2 Q_{\ell, m}(s, 2\Delta_{\CO}+2n)+O(N^{-6})\, .
}[]
In our setup, only the spin-zero tree-level anomalous dimension are non vanishing, we can therefore set $\ell=0$, and upon further specifying $\Delta_{\CO}=2$, we obtain
\eqna{
\CQ_{0, m}=-\frac{(\gamma_{n, \ell}^{(1)})^2}{N^4}\frac{ (2 n+3) \Gamma (2 n+3)^2 \Gamma (m+n+1)^2}{2 \Gamma (m+1) \Gamma (n+2)^4
   \Gamma (m+2 n+3)}+\cdots\,.
}[]
Plugging this into~\eqref{Mexp}, we find that at order  $N^{-2\kappa}$ the Mellin amplitude contains a contribution of the form
\eqna{
M_{(\kappa-1)\text{-loop}}(s,t)\supset \frac{-a_{0, \ell}^{(0)}(\gamma_{0, \ell}^{(1)})^\kappa}{(t-(4+2n+2m))^{\kappa-1}} \frac{ (2 n+3) \Gamma (2 n+3)^2 \Gamma (m+n+1)^2}{2 \Gamma (m+1) \Gamma (n+2)^4
   \Gamma (m+2 n+3)}\,.
}[]
This is precisely the term to be compared with $M^R_{(\kappa-1)\text{-loop}}$ in~\eqref{MR} in order to extract the residues $R_m^{(\kappa-1)}$.  After substituting the explicit OPE data and changing variables $m\to m-n$,  we finally obtain 
\eqna{
R_m^{(\kappa-1)}=-\sum_{n=0}^m \frac{4\cdot 3^\kappa (n+1)^{3\kappa}\Gamma(m+1)^2}{(2n+1)^{\kappa-1}(2n+3)^{\kappa-1}\Gamma(m-n+1)\Gamma(m+n+3)}\, .
}[]
Apart from the case $\kappa=2$ this expression cannot be resummed in an explicit closed form \footnote{It is however possible to get a recursive sequence.}. However, in the form given above it becomes particularly convenient to extract the large-$m$ behavior, which will be relevant in the following.  First observe that for $m,n$ large
\eqna{
\frac{\Gamma(m+1)^2}{\Gamma(m-n+1)\Gamma(m+n+3)}&=\frac{1}{(m+1)(m+2)}\frac{\begin{pmatrix}
2m+2\\
m-n
\end{pmatrix}}{\begin{pmatrix}
2m+2\\
m
\end{pmatrix}}\\
&\approx \frac{1}{(m+1)(m+2)}e^{-\frac{n(n+2)}{m+1}}\, ,
}[]
where we have used the expansion of the binomial coefficient $\begin{pmatrix}
x\\y
\end{pmatrix}$ for large argument.  Expanding then all the other terms, we obtain
\eqna{
R_m^{(\kappa-1)}&\approx \sum_{n=0}^m -\frac{3^\kappa n^{\kappa+2}}{4^{\kappa-2}m^2}e^{-\frac{n^2}{m}}\approx-\frac{3^\kappa m^\frac{\kappa-1}{2}}{4^{\kappa-2}} \int_{0}^\infty dx \lsp x^{\kappa+2}e^{-x^2}\\
&=-\frac{3^\kappa}{2^{2\kappa-3}}\Gamma\mleft(\frac{\kappa+3}{2}\mright)m^{\frac{\kappa-1}{2}}\, ,
}[]
where we have used $x=n / \sqrt{m}$.  Remarkably,  this expression makes manifest the connection between the residues at order $N^{-2\kappa}$ and the one-loop result
\eqna{
R_m^{(\kappa-1)}&\approx -6\lsp  \Gamma\mleft(\frac{\kappa+3}{2}\mright)\left(\frac{-2}{9\sqrt{\pi}}\right)^{\kappa-1} (R_m^{(1)})^{\kappa-1}\, .
}[largemR]
This factorized structure closely mirrors what happens in flat space, where higher-loop bubble diagrams can be understood as repeated insertions of the one-loop one. We will return to this analogy in more detail in Sec.~\ref{sec4}.
\subsection{Connection to $\hat{\gamma}_{0, \ell}^{(\kappa)}$}
So far, we have showed that the coefficients  $R^{(\kappa-1)}_m$  are uniquely determined from the tree-level OPE data. At the same time, in Sec.~\ref{sec2}, we have seen that the same information fixes the leading $\log J$ contributions to the anomalous dimensions at each loop order. This strongly suggests a direct correspondence between the $R_m^{(\kappa-1)}$ terms in~\eqref{MR} and the leading large-$J$ behavior of the anomalous dimensions in~\eqref{logJgen}.

As derived in detail in~\cite{Aharony:2016dwx},  by matching powers of $\log u$  and performing a conformal block decomposition in Mellin space (analogous to~\eqref{GCBl}), one can express the correction to the anomalous dimension at order $N^{-2\kappa}$ in terms of the corresponding Mellin amplitude
\eqna{
\gamma_{0, \ell>0}^{(\kappa)}=\frac{-1}{2\pi i} \int_{-i \infty}^{i\infty}\! \!\!\!ds \lsp M^\prime_{(\kappa-1)\text{-loop}}(s, 2\Delta_{\CO})\Gamma\mleft(\frac{s}{2}\mright)^2\Gamma\mleft(\frac{2\Delta_{\CO}-s}{2}\mright)^2\!{}_3F_2\left(\begin{matrix}-\ell,\,  \ell+2\Delta_{\CO}-1, \, \frac{s}{2}\\
\,\, \, \Delta_{\CO}, \, \,\,\, \Delta_{\CO}\end{matrix};1\right),
}[]
where $M^\prime_{(\kappa-1)\text{-loop}}(s, 2\Delta_{\CO})$ denotes the Mellin amplitude with the  $t=2\Delta_{\CO}$ pole subtracted.  For large $\ell$, or equivalently large conformal spin $J$,  the hypergeometric function admits the approximation
\eqna{
{}_3F_2\left(\begin{matrix}-\ell,\,  \ell+2\Delta_{\CO}-1, \, \frac{s}{2}\\
\,\, \, \Delta_{\CO}, \, \,\,\, \Delta_{\CO}\end{matrix};1\right)&\approx \frac{\Gamma(\Delta_{\CO})^2}{\Gamma\mleft(\frac{2\Delta_{\CO}-s}{2}\mright)^2}\beta_J(s)+(-1)^\ell\frac{\Gamma(\Delta_{\CO})^2}{\Gamma\mleft(\frac{s}{2}\mright)^2}\beta_J(2\Delta_{\CO}-s)\, , \\
\beta_J(s)&=\frac{1}{J^s}\left( 1+\frac{3\Delta_\CO s^2-s^3-3s^2-2s}{12J^2}+\cdots \right)\, .
}[]
At leading order this gives 
\eqna{
\gamma_{0, \ell \gg 1}^{(\kappa)}\approx -\frac{\Gamma(\Delta_{\CO})^2}{2\pi i} \int_{-i \infty}^{i\infty}\! \!\!\!ds \lsp M^\prime_{(\kappa-1)\text{-loop}}(s, 2\Delta_{\CO})\left( \frac{\Gamma\mleft( \frac{s}{2}\mright)^2}{J^s}+ \frac{\Gamma\mleft( \frac{2\Delta_{\CO}-s}{2}\mright)^2}{J^{2\Delta_{\CO}-2}}\right)\, ,
}[]
where we have set $(-1)^\ell=1$,  since only even spins are exchanged. From this expression, it is straightforward to see that the leading $\log J$ term can only arise from the part of the Mellin amplitude we have denoted $M^R_{(\kappa-1)\text{-loop}}$ in~\eqref{MR}.  
\eqna{
\gamma_{0, \ell \gg 1}^{(\kappa)}\approx -\frac{\Gamma(\Delta_{\CO})^2}{2\pi i} \int_{-i \infty}^{i\infty}\! \!\!\!ds \lsp \sum_{m=0}^\infty \frac{R^{(\kappa-1)}_m}{(s-(2\Delta_{\CO}+2m))^{\kappa-1}} \left( \frac{\Gamma\mleft( \frac{s}{2}\mright)^2}{J^s}+ \frac{\Gamma\mleft( \frac{2\Delta_{\CO}-s}{2}\mright)^2}{J^{2\Delta_{\CO}-2}}\right)\, ,
}[]
The first term can be evaluated by closing the contour to the right and picking up the pole at $s=2\Delta_{\CO}+2m$. Since each of them contributes a factor $J^{-(\Delta_{\CO}+2m)}$, for the leading contribution at large spin we can set $m=0$. To evaluate the second term, we can close the contour to the left and since there are no poles, it simply  vanishes.  We therefore obtain
\eqna{
\gamma_{0, \ell \gg 1}^{(\kappa)}&\approx\frac{(-1)^\kappa  }{\Gamma(\kappa-1) } \frac{ \log^{\kappa-2}\!J }{J^{2\Delta_{\CO}}}R^{(\kappa-1)}_0\Gamma(\Delta_{\CO})^4+\cdots\\
& =-a_{0,0}^{(0)}(\gamma_{0,0}^{(1)})^{\kappa} \frac{(-1)^\kappa \log^{\kappa-2}\! J }{2\Gamma(\kappa-1)}\Gamma(2\Delta_{\CO})\, ,
}[]
where in the second line we have used 
\eqna{
R_0^{(\kappa-1)}=-a_{0,0}^{(0)}(\gamma_{0,0}^{(1)})^{\kappa}\frac{\Gamma(2\Delta_{\CO})}{2\Gamma(\Delta_{\CO})^4}\,.
}[]
Notice that setting $\Delta_{\CO}=2$, this expression exactly reproduces out previous result in~\eqref{largeJscalar} and more generally, it makes manifest the direct correspondence between the leading $\log J$ behavior of the anomalous at $(\kappa-1)$-loops and  the restricted Mellin amplitude $M^R_{(\kappa-1)\text{-loop}}$.

\section{Flat-space limit}\label{sec4}
A holographic description of the flat-space S-matrix can be obtained by taking the AdS radius to infinity. In this limit, the bulk dynamics effectively reduce to those of flat spacetime, and correlators in the boundary CFT are expected to reproduce flat-space scattering amplitudes. As first emphasized in~\cite{Penedones:2010ue} and later further developed in~\cite{Fitzpatrick:2011ia}, this relation is most transparently realized at the level of the Mellin amplitude. Indeed the Mellin variables $s_{ij}$\footnote{With respect to our previous conventions: $s_{12}=t$, $s_{13}=s$ and $s_{14}=\hat{u}$.} admit a direct interpretation as the kinematic invariants in a scattering amplitude, $s_{ij}\sim p_i \cdot p_j$, and the flat-space amplitude can be obtained by taking them to be large. More concretely
\twoseqn{
M(s_{ij})&\approx \frac{R_{\rm AdS}^{-d+3}}{\CN \Gamma\mleft(\frac{1}{2}\sum_i \Delta_i-\frac{d}{2}\mright)}\int_{0}^\infty \!\!\!d\beta\lsp  \beta^{\frac{1}{2}\sum_i \Delta_i-\frac{d}{2}-1}e^{-\beta} A\left(S_{ij}=\frac{2\beta}{R^2_{\rm AdS}}s_{ij}\right)\,, \,  s_{ij}\gg 1\, , }[MfromA]
{A(S_{ij})&=\frac{\CN \Gamma\mleft(\frac{1}{2}\sum_i \Delta_i-\frac{d}{2}\mright)}{R_{\rm AdS}^{-d+3}} \!\!\lim_{R_{\rm AdS}\to \infty}\int_{-i \infty}^{+i \infty}\frac{d\alpha}{2\pi i }\alpha^{\frac{d}{2}-\frac{1}{2}\sum_i \Delta_i}e^\alpha  M\left(s_{ij}=\frac{R_{\rm AdS}^2}{2\alpha}S_{ij}\right)\, , 
}[AfromM][mellFlat]
where $R_{\rm AdS}$ is the radius of AdS$_{d+1}$,  $\CN$ is a normalization constant and $A(S_{ij})$ is the flat space scattering amplitude.

Let us illustrate this by analyzing the one-loop result.  For $\Delta_i=2$,  the Mellin amplitude takes the form
\eqna{
M_{\rm 1-loop}(s,t)=\sum_{m}\frac{R_m^{(1)}}{t-(4+2m)}+\text{ crossed}\, .
}[]
To apply the flat-space limit, we must take $s,t\gg 1$ large.  In this regime, the $m$  sum is dominated by terms with $m\sim s,t \gg 1$, so it can be approximated by an integral.  Focusing on a single channel, one finds
\eqna{
M_{\rm 1-loop}(t)\approx -\frac{27}{8}\sqrt{\pi } \int_{0}^\infty dm \frac{m^{1/2}}{t-2m}\, ,
}[]
where we have used the large $m$ approximation of $R_m^{(1)}$ in~\eqref{largemR}.  The integral is divergent,  but it can be regularized as
\eqna{
M_{\rm 1-loop}(t)=-\frac{27}{8}\sqrt{\pi} \int_{0}^\infty dm \frac{m^j e^{-w^2/m}}{t-2m}\, ,
}[]
so that once can first evaluate the integral and then set $j=\frac{1}{2}$ and $w=0$. This yields
\eqna{
M_{\rm 1-loop}(t) \stackrel{t\gg 1}{\approx} -\frac{27\pi^{\frac{3}{2}}\sqrt{-t}}{16 \sqrt{2}} \,. 
}[M1loop]
Plugging this result into~\eqref{AfromM} we get
\eqna{
A_{\rm 1-loop}=\frac{9\pi R_{\rm AdS}^2}{8N^4}\CN \sqrt{-T}\,.
}[]
This should be compared to the corresponding flat-space amplitude, namely the bubble diagram in five dimensions. A straightforward evaluation of the integral yields
\eqna{
A_{\rm 1-bubble}=\lambda_{\rm flat}^2\frac{\sqrt{-T}}{128 \pi}\, .
}[]
The two results have exactly the same $\sqrt{-T}$ behavior, and this comparison fixes the flat-space coupling to be
\eqna{
\lambda_{\rm flat}=-\frac{12 R_{\rm AdS}}{N^2}\sqrt{\CN}\, .
}[]
From two loops onward,  the flat space amplitude receives contributions from Feynman diagrams of different topologies, but  at each order there is always a contribution from a chain of $(\kappa-1)$ bubbles at $(\kappa-1)$-loop order. In flat space such diagrams are simple to compute, as they factorize:
\eqna{
A_{(\kappa-1)\text{-bubble}}=(A_{\rm 1-bubble})^{\kappa-1}\, .
}[]
Substituting this into~\eqref{MfromA} predicts a Mellin amplitude of the form
\eqna{
M_{(\kappa-1)\text{-bubble}}(t)&\approx \frac{1}{N^{2\kappa}} (-t)^{\frac{\kappa-1}{2}} \CN^{\frac{\kappa-2}{2}}  \pi  (-3)^{\kappa } 2^{\frac{13}{2}-\frac{9 \kappa }{2}} \Gamma \left(\frac{\kappa +3}{2}\right) \\
&=\frac{1}{N^{2\kappa}}(M_{\rm 1-loop}(t))^{\kappa-1} \CN^{\frac{\kappa-2}{2}}4 (-1)^{\kappa } 3^{3-2 \kappa } \pi ^{\frac{5}{2}-\frac{3 \kappa }{2}} \Gamma \left(\frac{\kappa
   +3}{2}\right)\, ,
}[kbubbles]
with $M_{\rm 1-loop}(t)$ in~\eqref{M1loop}.
It is now interesting to interpret this result in light of our expression for the $R_m$ coefficients.  Focusing again on a single channel, we recall that
\eqna{
M^R_{(\kappa-1)\text{-loop}}=\frac{1}{N^{2\kappa}}\sum_{m}\frac{R_m^{(\kappa-1)}}{(t-(4+2m))^{\kappa-1}}\, .
}[]
Using the large-$m$ expansion of the residues given in~\eqref{largemR}, this can be rewritten as 
\eqna{
M^R_{(\kappa-1)\text{-loop}}=-6\lsp  \Gamma\mleft(\frac{\kappa+3}{2}\mright)\left(\frac{-2}{9\sqrt{\pi}}\right)^{\kappa-1}\frac{1}{N^{2\kappa}}\sum_{m}\frac{ (R_m^{(1)})^{\kappa-1}}{(t-t_m)^{\kappa-1}}\, ,
}[largemR]
where we have defined $t_m\equiv 4+2m$.  This result should be compared to
\eqna{
(M_{\rm 1-loop}(t))^{\kappa-1}=\left( \sum_m \frac{R_m^{(1)}}{t-t_m}\right)^{\kappa-1}\,.
}[]
The two structures are evidently not equivalent: expanding the latter produces, in addition to the diagonal terms $\frac{ (R_m^{(1)})^{\kappa-1}}{(t-t_m)^{\kappa-1}}$, also cross terms with $m_i\neq m_j$.  However these cross-terms generate lower-order poles\footnote{As an example,  consider $\kappa=3$
\eqna{
(M_{\rm 1-loop}(t))^{2}&=\sum_m \frac{(R_m^{(1)})^2}{(t-t_m)^2}+2\sum_{m<n} \frac{R_m^{(1)} R_n^{(1)}}{(t-t_m)(t-t_n)}\, , \\
\frac{1}{(t-t_m)(t-t_n)}&=\frac{1}{(t_m-t_n)}\left( \frac{1}{t-t_m}- \frac{1}{t-t_n}\right)=\frac{1}{2(m-n)}\left( \frac{1}{t-t_m}- \frac{1}{t-t_n}\right)\,.
}[]}, thus contributing to subleading logarithms in position space. So as far as the leading log results is concerned we can approximate
\eqna{
M^R_{(\kappa-1)\text{-loop}}\Big|_{\rm leading\, log}=-6\lsp  \Gamma\mleft(\frac{\kappa+3}{2}\mright)\left(\frac{-2}{9\sqrt{\pi}}\right)^{\kappa-1}\frac{1}{N^{2\kappa}}(M_{\rm 1-loop}(t))^{\kappa-1}\, ,
}[]
and fixing $\CN=(2\pi)^2$ this exactly matches the expectation from $(\kappa-1)$-bubbles in~\eqref{kbubbles}. In this way, we have shown how $M^R$ precisely captures the contribution of bubble-like diagrams, providing the Mellin-space counterpart of the flat-space factorization property of bubble diagrams.

This correspondence with flat-space bubble diagrams also clarifies the role of the $R_m$ coefficients. In flat space, the $(\kappa-1)$-loop bubble diagrams are the only diagrams that admit a maximal cut in one channel, obtained by putting all propagators on shell. Concretely, each propagator is replaced by a delta function,
\eqna{
\frac{1}{p^2} \to 2\pi i \delta (p^2)\, ,
}[]
so that the diagram reduces to a product of on-shell phase-space measures. For a chain of $(\kappa-1)$ bubbles, this procedure yields
\eqna{
\text{max cut}_{t\text{-channel}} \left[ A_{(\kappa-1)\text{-loop}}\right]\propto (-T)^{\frac{\kappa-1}{2}}\, .
}[]
Remarkably, the same scaling is already encoded in the Mellin amplitude $M^R$, and in particular in the residues of its $(\kappa-1)$ poles. In the large-$m$ regime, it is enough to identify $m \simeq T/2$ in the expansion of $R_m$, obtaining
\eqna{
R^{(\kappa-1)}({m={T}/{2}}) \propto T^{\frac{\kappa-1}{2}} \, ,
}[]
which exactly reproduces the dependence of the flat-space maximal cut.
This observation, first emphasized in~\cite{Bissi:2020woe}, shows that the Mellin-space residues provide a direct counterpart of flat-space maximal cuts. More conceptually, it highlights that information contained purely in tree-level data already fixes universal aspects of loop amplitudes, as illustrated in Fig.~\ref{fig:maxcuts}.
\begin{figure}
\centering
\begin{tikzpicture}
\node at (0,0){$ \frac{1}{2^\kappa \kappa!}\log^\kappa\!  u\sum\limits_n u^{\Delta_{\CO}+n} \lsp  a_{n,0}^{(0)}\left(\gamma_{n,0}^{(1)}\right)^\kappa\tilde{g}_{2\Delta_{\CO}+2n}(u,v)\,\,  \longleftrightarrow\, \,   R_m^{(\kappa-1)}\, \,  \longleftrightarrow\quad $ };
\draw[thick] (6,0.3) circle (0.3);
\draw[thick] (6,0.9) circle (0.3);
\draw[thick] (6,-0.9) circle (0.3);
\filldraw[thick] (6,-0.15) circle (0.005);
\filldraw[thick] (6,-0.3) circle (0.005);
\filldraw[thick] (6,-0.45) circle (0.005);
\draw[thick] (6,1.2)--(6.2, 1.5);
\draw[thick] (6,1.2)--(5.8, 1.5);
\draw[thick] (6,-1.2)--(6.2, -1.5);
\draw[thick] (6,-1.2)--(5.8, -1.5);
\draw[red!70!black,dash pattern=on 1.5pt off 1pt, thick] (5.6,0.9)--(6.4, 0.9);
\draw[red!70!black,dash pattern=on 1.5pt off 1pt, thick] (5.6,0.3)--(6.4, 0.3);
\draw[red!70!black,dash pattern=on 1.5pt off 1pt, thick] (5.6,-0.9)--(6.4, -0.9);
\end{tikzpicture}\caption{Schematic relation between the order-$(\kappa-1)$ residues of the Mellin amplitude, fully determined by tree-level anomalous dimensions, and the maximal cut of the $(\kappa-1)$-loop bubble diagrams.}\label{fig:maxcuts}
\end{figure}
\section*{Acknowledgements}
We thank Alessandro Georgoudis for collaboration at early stages of this work.  GF thanks Liam Fitzpatrick and Wei Li for collaborations on related topics. 
A.B.\ and M.R.K.\ are partially supported by the  INFN Iniziativa Specifica ST\&FI. G.F.  is supported by the US Department of Energy Office of Science under Award Number DE-SC0015845, and is partially supported by the Simons Collaboration on the Non-perturbative Bootstrap.
\newpage
\appendix
\addtocontents{toc}{\protect\vskip1.5em}

\Bibliography[refs.bib]
\end{document}